# Energy and Power requirements for alteration of the refractive index

Jacob B Khurgin

Johns Hopkins University

Baltimore MD 21218 USA

**Abstract**

The ability to manipulate the refractive index is a fundamental principle underlying numerous photonic devices. Various techniques exist to modify the refractive index across diverse materials, making performance comparison far from straightforward. In evaluating these methods, power consumption emerges as a key performance characteristic, alongside bandwidth and footprint. Here I undertake a comprehensive comparison of the energy and power requirements for the most well-known index change schemes. The findings reveal that while the energy per volume for index change remains within the same order of magnitude across different techniques and materials, the power consumption required to achieve switching, 100% modulation, or 100% frequency conversion can differ significantly, spanning many orders of magnitude. As it turns out, the material used has less influence on power reduction than the specific resonant or traveling wave scheme employed to enhance the interaction time between light and matter. Though this work is not intended to serve as a design guide, it does establish the limitations and trade-offs involved in index modulation, thus providing valuable insights for photonics practitioners.

1. Introduction

Real time alteration of optical properties is at the heart of modern photonics, enabling many key devices such as modulators, switches, tunable filters, displays and many others[1]. Variation of the refractive index (the real part of permittivity) is obviously the most sought after, as it allows change of both phase and intensity (using various interference schemes) without incurring additional insertion loss[2]. Ideally, one is looking for the material capable of changing its refractive index by a large amount with low absorption and doing it at high speed with low power consumption. Unsurprisingly, these demands almost always contradict each other. Hence, trade-offs are inevitable, and, for each particular application, the characteristics central to performance must be defined. The quest for the "most suitable if not perfect" material with a variable index has been going on for decades if not longer and is expected to proceed in the foreseeable future, even though many spectacular successes have already been achieved, with new materials and devices emerging regularly. The natural question that arises is what (if any) fundamental limits exist for index variation? As the index can be varied by diverse means: temperature[3], electric[4] and magnetic[5] fields, electric charge[6], acoustic waves[7], strain[8], optical fields[9] , and more ,research has focused on determining limits for each specific mechanism and setup employed to modify the refractive index. Yet, a universal grasp of the nature-imposed limitations on index modulation, irrespective of the approach, is still lacking.

In this short essay I try to present a rather simple unified picture of index modulation and obtain an important characteristic parameter that is common for most solid materials – the energy density required to change the refractive index $n$ by roughly $\Delta n \sim n$ (one may say by about 100%). For the lack of a better definition, I shall refer to it as the VLIC ("Very Large Index Change") energy density. This VLIC energy density turns out to have a rather weak dependence on the modulation method and a strong dependence on the wavelength. But while this energy density is fundamentally important, from a practical point of view it is the power required to achieve 180 degrees of phase shift that determines performance of most devices. This power depends on speed, optical bandwidth, and footprint of the device in addition to the aforementioned VLIC energy density. All of it would require serious tradeoffs and compromises when choosing the material, modulation method, and particular implementation for each particular application that will be briefly described in this essay.

## 2. What makes optical range different?

Changing refractive index in the optical range is a challenging task. Typically, index changes are in the range of fractions of a percent, unless exorbitantly high-power densities are employed or the change in the real part of permittivity is accompanied by a large change in the imaginary part (i.e., excessive absorption). This is quite different from the dielectric properties in the low frequency domain, which for our purposes encompasses frequencies from DC to a few THz. In that range, the effective permittivity change is easy to achieve and is implemented routinely in many applications. One recent example is switching capacitors in a transmission line used to demonstrate time reflection at MHz frequencies[10]. Since the dielectric constant is nothing but the distributed capacitance of the medium, changing capacitance is tantamount to changing the dielectric constant and the speed of propagation of electromagnetic waves. Changing capacitance can be accomplished for instance by shunting a portion of the capacitor with a conductor, thus reducing the gap size and increasing the capacitance as schematically shown in Fig.1a, where short-circuiting two plates increases capacitance by a factor of 2. In a typical variable capacitor (varactor) [11], a change of capacitance is attained by depletion/accumulation of carriers in a reverse-biased *pn* junction, Fig.1b. However, the injected carriers cannot provide efficient conductive channels at frequencies $\omega \geq 1/RC$ and, for the simple geometry of Fig1b, the RC constant is equal to the dielectric relaxation time $\tau_d = \varepsilon_r \varepsilon_0 / \sigma$ where $\varepsilon_r$ is a dielectric constant and $\sigma$ is conductivity. The dielectric relaxation time is typically longer than 10's of femtoseconds even in highly doped semiconductors – hence at optical frequencies the injected carriers simply do not have time to react to electric field. In general, at high (optical) frequencies the admittance $i\omega C$ becomes so large that, in order to affect the effective impedance, one needs to either connect or disconnect capacitors with an extremely large conductance, such as can be provided by metals only [12]. That would require a mechanical switch that cannot be operated at a high modulation frequency. For this reason, the switched capacitor technique of permittivity change is inapplicable in the optical range, and one has to consider a different approach.

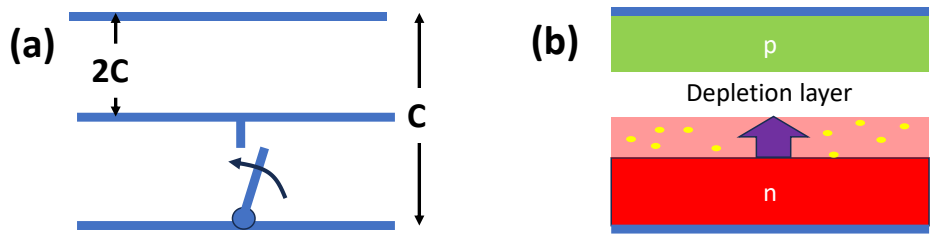

**Figure 1** (a) capacitance change by shorting the plates (b) practical implementation of a variable capacitor (varactor)

### 3. Model for very large index change (VLIC)

One shall only take one look at all the variety of the solid optical materials, transparent in, say the visible to the near IR region, to see that the indices of refraction range from 1.3 to maybe 4 at best[13]. If one takes a look at two very different materials with dissimilar crystal structures and bandgaps, say TiO$_2$ and diamond, their indexes vary by less than 10%[14] and the situation is even closer for say Si and GaAs[15]. It follows that changing the index by as much as 50% would require nothing less than removing one material and substituting another, very different one. Needless to say, unless this feat is performed literally, i.e., mechanically, an amount of energy commensurate with the cohesive energy holding the material would need to be supplied. Let us now test this assumption using a simple model.

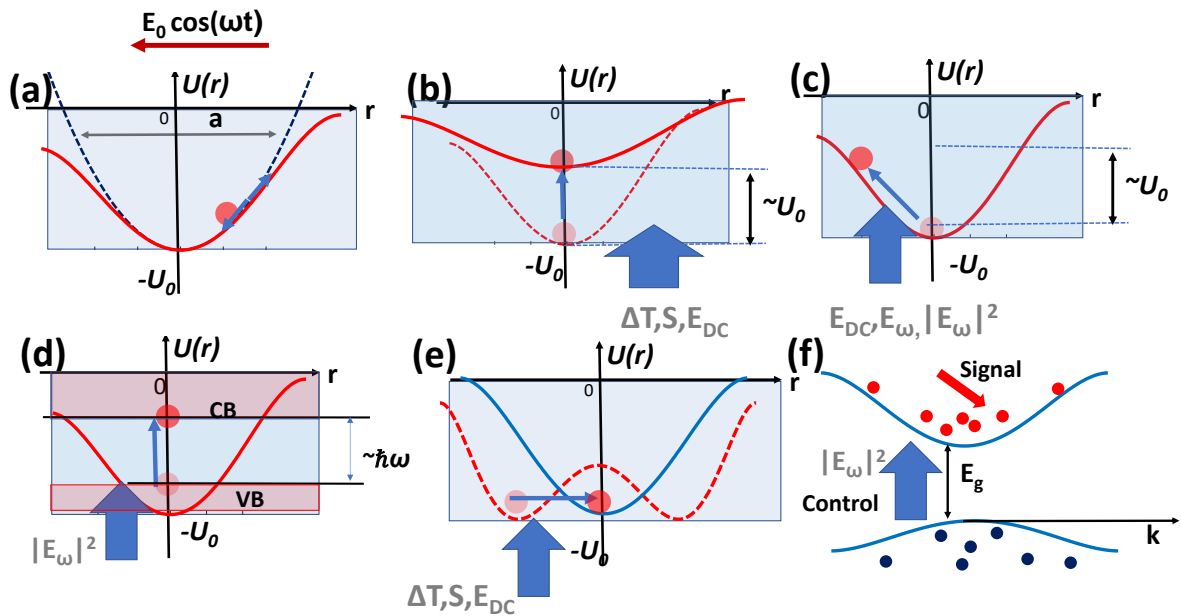

**Figure 2** A model for various means of imposing refractive index change: (a) a typical anharmonic potential of depth $U_0$ - the refractive index depends on curvature of $U(r)$; (b) the curvature can be changed by lattice distortion due to temperature, strain, or electric field; (c) the curvature can be changed by moving all electrons up within the potential due to electric field but staying in the same energy band (a.k.a. "virtual transition"); (d) Electrons can be moved into higher energy band ("real transition"), saturating the absorption and index via a Kramers-Kronig relation. (e) Near the phase transition, the shape of the potential changes significantly under a relatively small amount of change in temperature, field, or strain; the index change is enhanced at the expense of volatility. (f) Excitation of free carriers causes an index change due to their Drude-like response. Additional change is imposed by electrons moving to the higher states in the conduction band with larger effective mass.

In a most straightforward way, one can represent optical properties of the material by an anharmonic potential, $U(x) = -U_0 + Kx^2/2 - Lx^3/6 + ....$ as shown in Fig.2a. Then the equation of motion for an electron with energy $U_n$ subject to a harmonic field $E(t) \sim E_0 \cos \omega t$ is $m_0 d^2x/dt^2 = -dU_n/dx - eE(t)$, and non-resonant polarizability (assuming the frequency lies in the transparency region below all electronic resonances) is $\alpha_N \sim e^2/\varepsilon_0(d^2U_N/dx^2)$. The dielectric constant (electronic part of it) is then obtained by summation of individual polarizabilities, and averaging over all the electrons, one obtains

$$\varepsilon \sim 1 + \frac{Ne^2}{\varepsilon_0 \langle d^2U/dx^2 \rangle_N} \sim 1 + \frac{Ne^2}{\varepsilon_0 m_0 \bar{\omega}_0^2} \quad (1)$$

i.e., the off-resonant permittivity is represented by the Lorentz formula with mean resonant frequency $\bar{\omega}_0$. In solid material this frequency corresponds to the so-called Penn gap $E_P = \hbar \bar{\omega}_0$ which is typically larger than the optical bandgap energy by a few eV.[16]  Now, neglecting the vacuum permittivity, one can see that if one is to increase the permittivity by a factor of 2 (or down by a factor of ½), there are two ways to do it: (a) change the energy $U_N$ of each electron by the amount commensurate with $U_0$ and (b) change the electron density $N$ by the same factor of 2 or ½. One way this energy change can be accomplished is by changing the binding energy $U_0$ - essentially by inducing lattice distortion. That is what takes place when the material experiences a temperature rise[3] (thermal modulators), strain [17], (acousto-optic modulators (AO) as well as strain sensors[8]), or ion motion due to an electric field (ionic part of the electro-optic (EO) effect)[1] , all schematically shown in Fig.2b.  Alternatively, the electrons can be promoted along the x-axis  to higher energies due to the electric field of the low frequency (electronic part of EO effect) or by the optical field (optical nonlinearity of second or third order [9, 18]) as shown in Fig.2c. As long as the electron remains in the same electron band, one often refers to this motion as a "virtual transition" requiring no energy dissipation. The change of $N$ can be obviously accompanied by carrier injection or depletion[19-21] (current-driven modulators[22, 23] ) or by promotion of electrons from the state in the valence band to the conduction band[24-26] (index change occurs  due to  absorption saturation according to the Kramers-Kronig relation).  This is shown in Fig. 3d. Regardless of the approach employed, it is reasonable to hypothesize that the energy density essential for inducing a twofold increase or decrease in permittivity and index, referred to as the VLIC energy density is on the scale of $u_0 \sim NU_0$. What should be taken as $U_0$ is not 100% unambiguous, as one can think of the binding energy of the bond, the cohesive energy, the fundamental gap energy, the aforementioned Penn gap, or even ionization energy. Nevertheless, working within the framework of an order of magnitude estimate, it can be definitively stated that the value in question consistently remains within the range of a few eV. This value exhibits a clear tendency to rise in tandem with an increase in the fundamental bandgap.  For instance, taking the Penn gap as a measure, it ranges from 10eV in diamond to less than 3eV in PbSe and InSb[16]. That is easy to interpret using Fig.2a – as binding energy decreases, the potential well becomes more shallow, and the energy levels in it get closer to each other.   The density of the valence electrons in materials is typically on the scale of $10^{22} - 10^{23} cm^{-3}$; hence one can estimate the VLIC energy density to be in the range $u_0 \sim 10^3 - 10^5 J/cm^3$, with larger values corresponding to the materials with wider transmission region (higher bandgap). Obviously, adding this amount of energy in a small volume even for a short time interval may cause damage (if not evaporation). Nevertheless, this "universal" gauge is useful as it allows one to estimate the permittivity change when a smaller amount of energy is "injected" into the medium as $\Delta\varepsilon/\varepsilon \sim (\Delta u/u_0)^m$, where m=1 for the processes in which the permittivity change is proportional to the energy density (thermal, Kerr, and various $\chi^{(3)}$ effects), or m=1/2 for the Pockels, AO, or $\chi^{(2)}$ nonlinearity as discussed in subsequent sections. Before continuing, let us pause and ascertain whether this order of magnitude analysis actually makes any sense by comparing the above ad-hoc value of $u_0$ with the estimates for some of well-established modulation methods. The less patient readers can skip the next section and go directly to Table I where it is summarized that indeed the VLIC energy density for the diverse modulation mechanism falls with 1-100kJ/cm³ range, but anyone interested in the details, or is just enjoying an uninterrupted flow of thoughts and data can follow through the next section.

### 4. Does this model make sense?

*Thermal index change*

First, let us look at thermo-optical modulation in which thermally induced expansion causes bandgap shrinkage, i.e., a reduction of $U_0$ (Fig.2b). The index change is defined as $\Delta n = (dn/dT)\Delta T$ where $(dn/dT)$ is a thermo-optic coefficient and the temperature rise is $\Delta T = \Delta u/c_V \rho$, where $u$ is the energy density "injected" into the medium with specific heat $c_V$ and density $\rho$. Now, the VLIC energy density can be found as

$$u_0 = nc_V\rho/(dn/dT) \tag{2}$$

Using the example widely used in integrated photonics silicon [3] with $n=3.47$, $dn/dT = 1.8\times 10^{-4} K^{-1}$, and $c_V = 0.7 J/g\cdot K$ we obtain $u_0 = 3.1\times 10^4 J/cm^3$.

*Linear electro-optic effect*

Next, consider the Pockels effect[27] in which the index changes as $\Delta n = n^3 r_{ij} E_{RF}/2 = 1$, where $r_{ij}$ is the relevant component of the EO coefficient, and $E_{RF}$ is the applied low frequency electric field. A change of permittivity occurs due both to the ion motion causing lattice distortion (Fig.2b) and to the distortion of the electronic bond (electron moving in the potential well as in Fig.2c). The energy density introduced into the material is $\Delta u = \varepsilon_0 \varepsilon_s E_{RF}^2/2$, where $\varepsilon_s$ is the static dielectric constant. Then we obtain $u_0 = 2\varepsilon_0\varepsilon_s/n^4 r_{ij}^2$. Consider the workhorse of the EO modulator – lithium niobate[28] with $n=2.2$, $r_{33} = 30 pm/V$ and $\varepsilon_s = 28.7$. The energy required to achieve 100% index change is $u_0 = 2.4\times 10^4 J/cm^3$, i.e., well within the range obtain from our crude phenomenological model. While for many reasons (transparency range, damage threshold etc.) LiNbO₃ remains the material of choice, there are materials with higher Pockels coefficients, such as BaTiO₃ with $r_{42} \sim 900 pm/V$ [29, 30]. This large coefficient is owed to operating relatively close to the Curie temperature $T_c = 120°C$ at which ferroelectric-to-paraelectric phase transition takes place (For comparison, in lithium niobate $T_c = 1210°C$.). One can illustrate the large change of refractive index when the material goes from ferroelectric to paraelectric phase in Fig.2e. Near the Curie point, the ions are relative easy to move (the so-called soft phonon mode), but besides enabling a large Pockels coefficient, this facility of motion also underpins a very large static dielectric constant $\varepsilon_s \sim 2000$, so that $u_0 \sim 3\times 10^3 J/cm^3$, which is almost an order of magnitude less than in lithium niobate but is still within the range of phenomenological model. Then there are also EO organic polymers in which the Pockels coefficient can be on a scale of a hundred pm/V or more[31], which can bring the VLIC energy density to $10^3 J/cm^3$ or even less. That can be understood by the fact that polymers typically have a low melting point[32] i.e. the potential $U_0$ in Fig.2.a is small, but, hence increase in index change occurs at the expense of reduced stability, which so far prevented polymers from replacing lithium niobate in wide range of applications. The last thing that needs to be said is that the *photorefractive effect*[33] has the same VLIC energy density as EO effect density as modulation of index is caused by the electric field of trapped photoinduced carriers.

*Second order nonlinear optical effect*

The linear EO effect can be understood as the result of combining optical and low-frequency waves, essentially manifesting as a second-order nonlinear phenomenon. However, its treatment takes a distinct approach through employment of the Pockels coefficient, which characterizes alterations in impermeability. When both waves are in the optical range, one usually uses a different language that introduces the second order susceptibility $\chi^{(2)}$ via $P(\omega_1 \pm \omega_2) = \varepsilon_0 \chi^{(2)}(\omega_1 \pm \omega_2; \omega_1, \omega_2) E(\omega_1) E(\omega_2)$, where $P(\omega_1 \pm \omega_2)$ is the nonlinear polarization. But one can also interpret this relation as $P(\omega_1 \pm \omega_2) = \varepsilon_0 \Delta \varepsilon_r(\omega_2) E(\omega_1)$ where $\Delta \varepsilon_r(\omega_2) = \chi^{(2)} E(\omega_2)$ is the wave of permittivity modulated by the *control* wave at frequency $\omega_2$ from which the weaker *signal* wave at frequency $\omega_1$ is scattered, engendering sum and difference frequency waves. As shown in Fig.1c, when the amplitude of the electron oscillation in the asymmetric bond increases, the permittivity (and hence index) changes.

The oscillating Index change is $\Delta n(\omega_2) = \chi^{(2)} E(\omega_2)/2n$. Since the injected energy density is $\Delta u = \varepsilon_0 \varepsilon E(\omega_2)^2 / 2$, it follows that the VLIC energy density is $u_0 = 2\varepsilon_0 n^6 / |\chi^{(2)}|^2$. Now, using the example of GaAs[34] with one of the highest nonlinear susceptibilities $\chi^{(2)}_{14} = 340 pm/V$, and $n = 3.4$, the VLIC energy density for this process amounts to $u_0 = 2.3\times 10^5 J/cm^3$ - a high number, compared to the case of the linear EO effect. This difference is due to the two facts. Firstly, at optical frequencies, the ions do not participate in motion and do not contribute to index alteration. Secondly, different bonds in crystal structure have their nonlinear polarizabilities partially cancelled.

*Third order nonlinearity (optical Kerr effect)*

The index alteration in the case of the third order nonlinearity is similar to the case of the second order one, but being proportional to the square of optical field, the change can occur at either low or high frequency. For the third order nonlinearity, the relevant parameter is the nonlinear index $n_2$, and the index change is $\Delta n = n_2 I$ where $I = c\varepsilon_0 \varepsilon |E|^2 / 2n$ is a power density related to the energy density as $u = (c/n)I$. It follows then that the VLIC energy density can be estimated as $u_0 = n^2/cn_2$. Using silicon[35] ($n = 3.47$, $n_2 = 3.4\times 10^{-14} cm^2/W$) we obtain $u_0 = 1.2\times 10^4 J/cm^3$. In chalcogenide glasses such as GeSSb[36] the nonlinear index $n_2 = 3.8\times 10^{-14} cm^2/W$ is comparable to Si, while refractive index $n = 2.2$ is lower resulting in a lower VLIC energy density $u_0 = 4.2\times 10^3 J/cm^3$. Another common material, silica glass[37] has nonlinear index that is two order of magnitude less than Si, $n_2 \sim 2\times 10^{-16} cm^2/W$ leading to very large values of VLIC energy density of more than $u_0 = 10^6 J/cm^3$, but due to the advantages of long propagation length in optical fiber, it is still widely used in practice. The impact of propagation length (traveling wave enhancement) will be explored further in the subsequent sections.

*Free carrier induced index change*

An altogether different way of changing the refractive index is based on modulation of the carrier density or the effective mass in doped semiconductors[38-40]. The dielectric constant is described by the Drude formula

$$\varepsilon_r(\omega) = \varepsilon_\infty - \frac{N_c e^2}{\varepsilon_0 m_c (\omega^2 + i\omega\gamma)} \quad (3)$$

where $N_c$ is the carrier density in the band, $\gamma$ is the damping rate, and $m_c$ is the effective mass. When the carrier density is photoinduced by the strong optical pump (control) at frequency $\omega_c$, the dielectric constant at signal frequency $\omega$ described by (3) changes. Clearly, the VLIC at frequency $\omega$ is achieved when $N_c = \varepsilon_\infty \varepsilon_0 m_c \omega^2/e^2$. Accordingly, the VLIC energy density is $u_0 = N_c \hbar \omega_c = \varepsilon_\infty m_c \omega^2 \omega_c / 4\pi\alpha_0 c$, where $\alpha_0$ is a fine structure constant. The situation is schematically represented in Fig.2.f. Clearly, one gets a certain benefit from the fact that the effective mass is small, and, especially from the fact that the operating frequency is also low, typically in the near IR range. Using the example of ZnO [41] $\varepsilon_\infty = 4$, $m_c = 0.25$, pump wavelength of $256 nm$, and probe at $1550 nm$, one obtains a photogenerated carrier density of $N_c \sim 10^{21} cm^{-3}$ and $u_0 \sim 800 J/cm^3$, on the lower end of the VLIC energy densities predicted phenomenologically (this mostly can be explained by the small effective mass). But that change is strongly frequency dependent and always accompanied by a strong absorption of the signal light.

When the incoming photon energy is less than the bandgap energy, the intraband absorption causes heating of the carriers in the non-parabolic band and the ensuing increase of the effective mass with a commensurate increase of permittivity[40, 42, 43]. The effective mass changes by roughly 50% when the energy of a carrier in the band becomes commensurate with the Bandgap energy; hence the VLIC energy density for intraband excitation is comparable to that for the interband case (although the sign of the index change is opposite) and the main difference lies in the time scale – as interband processes occur on a nanosecond scale (recombination time) while intraband on sub-picosecond times (electron-phonons scattering time).

Refractive index alteration also occurs at frequencies below the bandgap. As electrons transition to the conduction band, absorption reaches saturation, causing the absorption edge to shift upward. This shift leads to changes in the refractive index, consistent with the Kramers-Kronig relation, as depicted in Figure 2d. Achieving VLIC requires the excitation of a significant fraction of electrons, and the VLIC energy density of is determined by the equation $u_0 \sim NE_{gap}$. Although there's a possibility of decreasing this energy requirement by utilizing exciton resonance resonance[24], it's important to note that intrinsic material resonances, while effective for absorption modulation, demonstrate limited efficacy for index modulation purposes.

The change of refractive index can also be induced electrically, via injection in a *pn* junction[20] or in an MOS capacitor [6, 19]. The energy density required to attain the change is obviously $u \sim eN_c V$ where the voltage $V$ applied across the capacitance is typically commensurate with the bandgap which makes the VLIC energy density typically on the order of $1,000 J/cm^3$, commensurate with the all-optical modulation.

*Strain induced index change and acousto-optic effect*

The refractive index change due to strain $S$ is caused by lattice distortion (Fig.2b) and is determined by photo-elastic coefficient $P$ [44]as $\Delta n = n^3 PS/2 = 1$, while the energy density is $u = YS^2/2$ where $Y = \rho v_s^2$ is Young's modulus, and $v_s$ is the speed of sound. Then the VLIC energy density is $u_0 = 2n^2/M_2 v_s$, where $M_2 = n^6 P^2 / \rho v_s^3$ is the AO figure of merit. Consider now a widely used material (in AO modulators) TeO$_2$ [45] with $n = 2.1$, $M_2 = 35 \times 10^{-15} s^3/kg$, and $v_s = 4,250 m/s$. Its VLIC energy density is $u_0 = 6.1 \times 10^4 J/cm^3$. For the sake of completeness, we can also consider index change in the optical fiber used in strain sensors, where the experimentally measured relative change of index is linear with strain $\Delta n/n \sim KS$. Here, typically [46] $K \sim 0.75$ and the VLIC energy density can be found as $u_0 = nY/2K^2$. With $Y \sim 70 GPa$, one obtains $u_0 = 9.4 \times 10^4 J/cm^3$.

With that, we summarize the results in table I which confirms that the range of VLIC energy density is 1-100kJ/cm$^3$ for all the investigated mechanisms of index change.

| Modulation Method | Index change. Energy density | Expression for $u_0$ | Material | Material parameters | Value of $u_0$ (J/cm$^3$) |
|---|---|---|---|---|---|
| **Thermal** | $\Delta n = (dn/dT)\Delta T$ <br> $\Delta u = c_v \rho \Delta T$ | $u_0 = nc_v \rho/(dn/dT)$ | Si[3] | $n = 3.47$, <br> $\rho = 2.3 g/cm^3$ <br> $c_V = 0.7 J/g \cdot K$ <br> $dn/dT = 1.8 \times 10^{-4} K^{-1}$ | $3.1 \times 10^4$ |
| **Electro-optic** | $\Delta n = n^3 r_{33} E_{DC}/2$ <br> $\Delta u = \varepsilon_0 \varepsilon E_{DC}^2/2$ | $u_0 = 2\varepsilon_0 \varepsilon_s / n^4 r_{33}^2$ | LiNbO$_3$[28] | $n = 2.2$, $\varepsilon_s = 28.7$ <br> $r_{33} = 30 pm/V$ | $2.4 \times 10^4$ |
| | | | BaTiO$_3$[29] | $n = 2.2$ $\varepsilon_s \sim 2000$ <br> $r_{42} \sim 900 pm/V$ | $3 \times 10^3$ |
| **Optical Kerr** | $\Delta n = n_2 I$ <br> $\Delta u = In/c$ | $u_0 = n^2/cn_2$ | Si[35] | $n_2 = 3.4 \times 10^{-14} cm^2/W$ | $1.2 \times 10^4$ |
| | | | GeSSb[36] | $n_2 = 3.8 \times 10^{-14} cm^2/W$ | $4.2 \times 10^3$ |
| **2$^{nd}$ order nonlinear** | $\Delta n = \chi^{(2)} E_\omega / 2n$ <br> $\Delta u = \varepsilon_0 n^2 E_\omega^2/2$ | $u_0 = \varepsilon_0 n^6/2|\chi^{(2)}|^2$ | GaAs[47] | $n = 3.4$, <br> $\chi^{(2)} \sim 200 pm/V$ | $23.1 \times 10^4$ |
| **Acousto-optic** | $\Delta n = n^3 PS/2$ <br> $\Delta u = \rho v_s^2 S^2/2$ | $u_0 = 2n^2/M_2 v_s$ | TeO$_2$[45] | $n = 2.1$ <br> $M_2 = 35 \times 10^{-15} s^3/kg$ <br> $v_s = 4,250 m/s$ | $6.9 \times 10^4$ |
| **Strain** | $\Delta n/n \sim KS$ <br> $\Delta u = YS^2/2$ | $u_0 = nY/2K^2$ | Silica fiber[46] | $n = 2.1$, $K = 0.75$ <br> $Y \sim 70 GPa$ | $9.4 \times 10^4$ |

| Free carriers | $\Delta n/n \sim \Delta\omega_p/\omega_p$  $\Delta u \sim N_c \hbar\omega_{pump}$ | $u_0 \sim N_c \hbar\omega_{pump}$ | ZnO[41] | $N_c \sim 10^{21} cm^{-3}$  $\hbar\omega_{pump} \sim 4.8 eV$ | $0.8 \times 10^3$ |

**Table I** VLIC energy density for various index modulation techniques

### 5. Liquid crystals and opto-mechanics

Prior to progressing further, it is prudent to revisit the method of "mechanical" index modulation, wherein the straightforward substitution of a high-index material with a low-index counterpart, preferably air, is achieved through mechanical manipulation. The act of material substitution inherently proves to be a more feasible undertaking than the substantial alteration of its inherent properties. One might seek insight from historical alchemical practices or consider an example of turning graphite into diamonds which requires a great deal of pressure and compare it with speedily swapping one material for another that requires far lesser effort, as any adept conjurer can attest.

One way in which mechanical motion can be used to attain VLIC is by rotating birefringent entities, which is exactly what is done in liquid crystal (LC) modulators [48] where the individual molecules are rotated to align with applied electric field, changing index from ordinary to extraordinary. The electric fields required for rotation are relatively small (on the order of $10^7 - 10^8 V/m$) causing the VLIC energy density to be less than $1 J/cm^3$, but the response time of LC modulators is larger than 100ns which limits their use in many applications[49].

Similarly, optomechanical modulators with electro-statically activated mirrors [50] can be considered belonging to the same class. An ideal mirror can be thought of as a dielectric with infinitely high permittivity. But the operational bandwidth of these modulators is quite limited by inertia as accelerations involved in fast motion become unsustainable.

It is fair to say then that when high modulation speed is not required, optomechanical (digital micromirror) or LC phase modulators require the least energy and that is why they are widely used for such applications as displays[51] or in spatial light modulators[52, 53], especially in adaptive optics. Unfortunately, these modulators are not easily adaptable to photonic integrated circuits and other waveguide geometries, and one is often forced to use less energy-efficient index modulation methods, such as thermo-optical to enact relatively slow index change.

From this point in our discourse on, we shall focus only on the modulation methods that do not involve any mechanical motion and thus can in principle be operated with GHz and higher bandwidth. And, with a high degree of confidence we can say that to attain relatively fast large index change the energy density on the aforementioned scale of 1-100 kJ/cm³ must be introduced somehow into the material depending on the operational wavelength (higher for visible and UV and less for IR ranges). But what matters is not the amount of energy introduced (and this energy may be dissipated or recycled) but the operational power required to attain this change, and modulating schemes with similar VLIC energy densities may deviate substantially in terms of operational power. Furthermore, with the exception of recently popular time reflection and photonic time crystal ideas [54, 55], VLIC is not in fact required and all one needs is the phase change sufficient for optical switching (or 100% modulation depth) which is typically achieved with lesser index change and longer propagation distances. Nevertheless, the concept of VLIC energy density introduced here can serve a key role in allowing direct comparison of diverse modulation schemes, to which the rest of this work is devoted.

## 6. A practical switching criterion

The literature abounds with claims of substantial index changes, accompanied by impressive coefficients characterizing nonlinear and EO properties. However, a pivotal aspect often omitted from this discourse is whether these substantial changes genuinely translate into the efficiency of practical devices. The enduring prominence of lithium niobate in EO and frequency conversion applications, and the prevalent use of silicon and SiN for harnessing the optical Kerr effect indicate that perhaps VLIC by itself is not enough. A comprehensive survey of active optical devices—ranging from modulators and couplers to frequency converters—reveals a decisive criterion for achieving optimal performance. In the context of full switching in digital devices or achieving 100% modulation depth in analog devices, the phase shift engendered by the index change $\Delta\Phi = \Delta n(\Omega) k_0 L$, where $\Omega$ is the frequency of control (modulating) input, $k_0 = \omega/c$ is the signal wave vector, and the length $L$ is plays the key role. Mathematically, this phase shift should be on the order of $\pi/2$

This fundamental principle applies regardless of the modulation frequency $\Omega$ —whether it's low for EO and AO devices or high for all-optical frequency converting devices. This fact follows from the most fundamental considerations. By changing the refractive index one changes the photon momentum as $\Delta p = \Delta n \hbar k$ and, according to the uncertainty principle, in order to fully register this change (i.e. to measure index with precision exceeding $\Delta n$, the measuring apparatus (i.e. some type of interferometer, coupler, or frequency discriminator) must have length defined by $\Delta p L \geq \hbar/2$. Then $\Delta n k_0 L \geq 1/2$. The lower limit of ½ corresponds to a Gaussian distribution of index change and, for the case when the modulation region is uniform, it is reasonable that $\Delta\Phi = \pi/2$ may serve as a good estimate of full modulation.

This criterion of full modulation remains remarkably consistent across various scenarios. When index change is modulated in space—typical of AO modulators or grating-assisted couplers—or when index modulation occurs at optical frequencies, as in frequency converting devices, full conversion mandates an analogous criterion $\Delta n(\omega) k_0 L \geq \pi/2$,

Introducing the transit time $\tau_{tr} = nL/c$ the criterion for full switching/frequency conversion can be written as $\Delta n \omega \tau_{tr} > n\pi/2$, and since phase shift gets enhanced in the so called slow light schemes by a factor $c/nv_g$, where $v_g$ is the group velocity [56, 57], this expression $\Delta n \omega \tau_{tr} > n\pi/2$ is still valid as long as transit time is defined as $\tau_{tr} = L/v_g$. It should be noted that in often discussed "epsilon near zero"(ENZ) materials [39, 40] enhancement of phase shift can be always traced to the reduction of group velocity[58] ,thus putting ENZ materials into the "slow light" class. One can also consider some kind of resonant structure where the propagation path is effectively folded (this can be a ring resonator, a Fabry-Perot cavity, or a more complicated structure such as a photonic crystal, metasurfaces or a bound state in continuum). In such a case, the full switching/conversion criterion becomes $\Delta n \omega \tau_p > n\pi/2$, where $\tau_p$ is the photon lifetime in the resonator.

## 7. Linear vs. Quadratic modulation

Let us now see how the concept of VLIC energy density can be applied to obtain the expected value of operational power for the diverse modulation methods. First, consider the class of phenomena where the energy density is proportional to the square of the "force" causing the index change, i.e., strain (AO effect) or electric field (Pockels or second optical nonlinearity). The index change is then $\Delta n / n \sim (\Delta u / u_0)^{1/2}$ and the energy density needed to achieve full switching is

$$u_{sw}^{(1/2)} = \left[\frac{\pi}{2nk_0 L}\right]^2 u_0 \tag{4}$$

If, on the other hand the energy density is proportional to the square of the "force" as in light intensity (Optcal Kerr effect), Temperature (thermooptical), or optically induced carrier population, it follows that $\Delta n / n \sim \Delta u / u_0$, and

$$u_{sw}^{(1)} = \left[\frac{\pi}{2nk_0 L}\right] u_0 \sim 2nk_0 L u_{sw}^{(1/2)} \tag{5}$$

Since the parameter $Q = nk_0 L = \omega \tau_{tr}$, where $\tau_{tr} = nL/c$ is a transit time, can easily reach 100's and 1000's due to either a realistically long propagation length or a realistically high Q resonator, it is obvious that EO and AO modulators can operate with powers that are substantially lower than all-optical phase modulators based on Kerr effect or free carrier effects. The reasons that modulators other than Pockels and AO are used have little to do with power consumption – usually it is just size constraints and difficulties with velocity matching, as well as engineering electrodes and transducers.

## 8. Power requirements in "slow" and "fast" modulation methods

All the modulation schemes considered by us can be broadly divided into two classes according to whether the switching energy $u_{sw}$ introduced into the medium gets dissipated inside the medium or not[42].

*"Slow modulation"*

The first class includes all the schemes where index change is associated with energy dissipation e.g. optically driven free carrier modulation[40, 41, 43, 59], modulation based on absorption saturation, thermal nonlinearity, and photo-refractive effect among others. Each of these processes is limited by a characteristic relaxation time $\tau_r$ - which can be the recombination time, heat diffusion time, hot carrier thermalization time, and so on. This time defines the maximum operational frequency of modulator, $\Omega_{max} \sim 1/\tau_r$, as well as switching power,

$$P_{sw} = u_{sw} S_{eff} L / \tau_r = \frac{\pi}{2n_g} \frac{1}{\omega \tau_r} c u_0 S_{eff} = \frac{\pi}{2n_g} \frac{\Omega_{max}}{\omega} c u_0 S_{eff}, \tag{6}$$

where $S_{eff}$ is the effective cross-section of the modulator[22, 23] that incorporates the confinement factor (i.e. overlap between the control and signal waves and active material). The power for a given material does not depend on operational frequency. Therefore, it makes sense to operate at maximum frequency, $\Omega \sim \Omega_{max} \sim 1/\tau_r$ i.e., to adjust the relaxation time to the required operational frequency, which in practice amounts to choosing a material with a suitable relaxation time. In some cases, the relaxation time itself can be "tuned" by incorporating defects [60] or designing intersubband quantum well structures [61]. Note, although we refer to these modulation schemes as "slow", the relaxation time can be quite short, less than a picosecond in case of intraband free carrier modulation[42], intersubband modulators[61], or interband modulation with a large number of defects reducing the recombination time[60]. Note that the switching power does not depend on the length of the modulator. Assuming a photonic waveguide with $S_{eff} \sim 0.1 \mu m^2$, $\lambda = 1550 nm$ and 100GHz operational frequency one can obtain switching power of a few watts (switching energy per bit of a few picojoules). These powers are way too high – hence plasmonic[62, 63] (or hybrid [6, 19] waveguides with $S_{eff} \sim 0.001 \mu m^2$) must be used to bring the switching power to a few milliwatts and the switching energy per bit to a few femtojoules. Regrettably, due to the inherent nature of high loss[64], plasmonic waveguides have faced limitations in their commercial utilization up to the present date.

*"Fast" modulator*

Next consider the "fast" modulator – fast in a sense that no energy dissipation is required to change the index of refraction. That, of course, does not mean that dissipation does not occur in the fast modulators. There can be a significant microwave loss in the electrodes in EO modulators, sound wave absorption in AO modulators, and parasitic absorption in Kerr modulators. But energy dissipation is not related to index change and thus is not essential for the operation and can be considered a "collateral damage". The speed of the fast modulator is determined either by external circuit parameters or by geometry, or it can be essentially unlimited, as in the case of nonlinear optics well below the absorption edge. First, we turn our attention to the lumped modulator in which the incoming control energy does not propagate along with the optical signal wave. Then we obtain

$$P_{sw} = u_{sw} S_{eff} L\Omega = \begin{cases} \dfrac{\pi^2}{4 n_g^2} \dfrac{\Omega}{\omega} \dfrac{cu_0}{k_0 L} S_{eff} & m = 1/2 \\ \dfrac{\pi}{2 n_g} \dfrac{\Omega}{\omega} cu_0 S_{eff} & m = 1 \end{cases} \quad (7)$$

where $m = 1/2$ corresponds to the EO effect, the AO effect, and the second order optical nonlinearity, while $m = 1$ is for the Optical Kerr effect and other third order effects based on it (including four wave mixing, cross-phase modulation and so on) . The key difference from the "slow" modulator is that switching power depends only on frequency, and operating at slower speed generally does reduce the power consumption commensurably. For the "lumped" optical Kerr effect, one obtains similar power requirements as for the slow modulator, but, for the lumped EO or AO modulator, the maximum length is limited to $L_{max} = \pi c / n_g \Omega$ (propagation delay should be less than half the period of the modulation wave). Hence

$$P_{sw,\min} = \frac{\pi}{4n_g} \frac{\Omega^2}{\omega^2} u_0 c S_{eff} \qquad (8)$$

Clearly, for 100GHz operation, the power in the "fast" modulator in which the index change is proportional to the square root of energy density, the switching power is reduced by a factor of $\Omega/\omega$ i.e., by up to three orders of magnitude or a few mW. This reduction is due simply to the propagation factor $k_0 L = \omega/\Omega$ that relates the switching energies in m=1 and m=1/2 modulators.

### 9. Resonant and traveling wave enhancements

*Resonant enhancement of the control wave*

Since the index change in "fast" modulators increases with "stored" and not dissipated energy, power requirements may be lowered significantly by employing resonances to enhance stored energy. In case of EO or AO, one can use a resonant LC circuit, a microwave or acoustic resonator with a given Q-factor $Q = \Omega \tau_{el(ac)}$, where $\tau_{el(ac)}$ is the lifetime of the electric (acoustic) excitation in the resonator. Then one obtains

$$P_{sw,\min} = \frac{\pi}{4n_g} \frac{\Omega}{\omega^2 \tau_{el(ac)}} u_0 c S_{eff} \qquad (9)$$

But since the signal bandwidth is now limited is $B \sim 1/2\pi \tau_{el(ac)}$, the expression for minimum switching energy becomes

$$P_{sw,\min} = \frac{\pi^2}{4n_g} \frac{\Omega B}{\omega^2} u_0 c S_{eff} \qquad (10)$$

In other words, for digital modulation $\Omega \sim B$ using a resonant circuit or a microwave resonator to enhance the control field hardly reduces power consumption, but for narrow band analog modulation on top of a high frequency microwave carrier (which is common in microwave photonics applications[65, 66]) there are certain advantages. When it comes to Kerr modulation one obtains

$$P_{sw} = \frac{\pi}{2n_g} \frac{1}{\omega \tau_p} c u_0 S_{eff} = \frac{\pi}{2n_g} \frac{B}{\omega} c u_0 S_{eff} \qquad (11)$$

where $\tau_p$ is a photon lifetime. This expression, however, assumes that only the modulating wave is enhanced by the resonator, while in reality both modulation and signal waves are enhanced as can be seen later in this section

*Traveling wave enhancement*

Next, we consider traveling wave enhancement. It is possible for of EO and collinear AO[67] modulation, or all-optical modulation and frequency conversion – in these devices one can achieve velocity and phase matching between the control wave and propagating optical signal. The switching power is related to the energy density as

$$P_{sw} = cu_{sw}S_{eff}/n_g = \begin{cases} \dfrac{\pi^2}{4n_g^3}\dfrac{u_0}{k_0^2L^2}cS_{eff} & m=1/2 \\ \dfrac{\pi}{2n_g^2}\dfrac{1}{k_0L}cu_0S_{eff} & m=1 \end{cases} \quad (12)$$

In comparison to the lumped modulator (7) one obtains a travelling wave enhancement of $Q_{tr} = Ln_g\Omega/c = \Omega\tau_{tr}$, where $\tau_{tr} = Ln_g/c$ is the transit time. One can also express the enhancement as $Q_{tr} = 2\pi Ln_g/\Lambda$ which indicates that in traveling wave geometry one need only supply power to the first $\Lambda/2\pi n_g$ segment of the waveguide (where $\Lambda$ is the wavelength of control wave). Since transit time does not affect the frequency and bandwidth of the modulation (it only affects latency), this factor can be quite large, with transit time (i.e., length) limited by the practical considerations of small footprint, or by loss. Overall, traveling wave modulation does offer the lowest switching power – for 1cm long modulator it can be less than $100\mu W$ in a waveguide configuration (assuming that good overlap between optical signal and modulating waves can be achieved). It also explains why the optical fiber, where the VLIC energy density $u_0$ is very large, is still most widely used medium for demonstrating all optical switching, frequency conversions, and more[68].

*Resonance at the optical signal frequency*

Alternatively, if an optical resonator centered at the optical signal frequency $\omega$ is used, then one can introduce the finesse of the resonator as $F = \pi c\tau_p/Ln_g = \tau_p/\tau_t$, where $\tau_p$ is photon lifetime, and $Ln_g/c = \tau_{tr}$ is a transit time. In a ring resonator the phase shift is enhanced by this very factor $F$ and then we obtain for EO or AO modulation

$$P_{sw} = \frac{\Omega}{\omega}\frac{cu_0}{k_0L}\frac{S_{eff}}{F^2} = \frac{\pi}{4n_g}\frac{\Omega}{\omega^2\tau_p}cu_0\frac{S_{eff}}{F} \quad (13)$$

For the case of Kerr modulation both modulating and signal waves are enhanced and we obtain

$$P_{sw} = \frac{\pi}{2n_g}\frac{1}{\omega\tau_p F}cu_0 S_{eff} \quad (14)$$

Note that both finesse and Q-factor $Q = \omega\tau_p$ are present in (13) and (14). It is important to recognize that there are numerous methods to increase the lifetime of photons. These methods range from basic ring[69, 70] and Fabry-Perot[71, 72] resonators to more advanced techniques involving photonic crystals[73], bound states in a continuum[74, 75], Fano resonances[76], and different metasurfaces[77]. Despite the complexity of these techniques, they all share the common aspect of being defined by the photon lifetime and Q factor. These sophisticated structures, however, don't offer significant improvements over a simple ring structure. Therefore, the decision to use them or not should be based on how well they align with the specific task and geometry, rather than assuming they inherently provide remarkable enhancements

## 10. Discussion

The analysis presented in this study can be summarized as follows: The power required to achieve switching, 100% modulation, frequency conversion, etc., using various index modulation methods can be represented as

$$P_{sw} = P_0 / X, \tag{15}$$

where $P_0 = c u_0 S_{eff}$ represents "basic power flow" associated with the intrinsic properties of the medium where modulation occurs and the waveguide cross-section, and where $X$ denotes the power reduction factor, which is linked to the interaction time between the medium and the optical wave. For photonic waveguides, the basic power typically falls within the range of $P_{0,ph} \sim 10^4 - 10^6 W$, depending on the operational wavelength, while in plasmonic and hybrid waveguides it can be 2-3 orders of magnitude lower $P_{0,pl} \sim 10^2 - 10^4 W$, albeit at the cost of higher insertion loss. Remarkably, the fact that $P_0$ does not change significantly for different materials and modulation methods can be explained using Feynman's interpretation[78] of quantum mechanics, wherein the strength of a phenomenon depends solely on the value of the matrix element of the interaction Hamiltonian and the interaction time. In the context of optical and dielectric properties, the interaction Hamiltonian is always determined by the dipoles, which remain on the same scale of a few Angstroms (more for IR and less for visible and UV). Thus, the variation in interaction times becomes the primary factor that can be adjusted by many orders of magnitude[79].

To compare power reduction schemes, we introduce the modulation period $T = 2\pi / \Omega$. For simplicity, we omit factors that are on the order of unity, and the power reduction factors are listed in Table 2. In the left column we show the schematics of modulation with the signal optical wave shown in red and the control input shown in blue.

| | Modulator | slow | Fast m=1/2 | Fast m=1 |
|---|---|---|---|---|
| 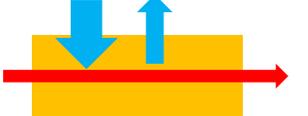 | Lumped | $X = \omega \tau_r$ $X_{max} = (\omega T)$ | $X = \omega \tau_{tr} \times \omega T$ $X_{max} = (\omega T)^2$ | $X = \omega T$ $X_{max} = (\omega T)$ |
| 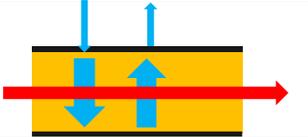 | Resonant control | | $X = \omega \tau_{tr} \times \omega \tau_{el(ac)}$ $X_{max} = (\omega T)^2$ | $X = \omega \tau_p$ $X_{max} = (\omega T)$ |
| 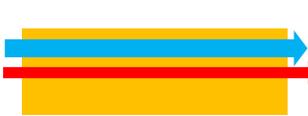 | Traveling wave | | $X = (\omega \tau_{tr})^2$ | $\omega \tau_{tr}$ |
| 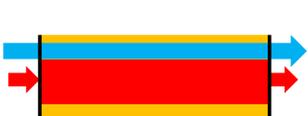 | Resonant signal | $X = \omega \tau_r \times \tau_p / \tau_{tr}$ $X_{max} = (\omega T) T / \tau_{tr}$ | $X = \omega T \times \omega \tau_p \times \tau_p / \tau_{tr}$ $X_{max} = (\omega T)^2 T / \tau_{tr}$ | $X = \omega \tau_p \times \tau_p / \tau_{tr}$ $X_{max} = (\omega T) T / \tau_{tr}$ |

**Table 2** Enhancement of Phase modulation (reduction of switching power) in various modulator configurations shown schematically in the first column. The control input is shown in blue and the signal wave in red.

As observed, the power reduction factors predominantly take the form of an "effective interaction time factor," $\omega\tau_{eff}$ or a combination thereof. The "effective interaction time" can represent the relaxation (recombination, thermalization, diffusion etc.) time $\tau_r$ in "slow" materials where index change requires energy dissipation. Alternatively, it can denote the period of modulating (control) wave $T$ over which the cumulative effect of index change occurs. Moreover, it might correspond to the propagation (transit) time $\tau_{tr}$, the lifetime of excitation in a microwave or acoustic resonator $\tau_{el(ac)}$, or even the photon lifetime $\tau_p$ in a resonator. The physical mechanisms enabling power reduction can vary, encompassing intrinsic and extrinsic factors. However, in the end, the key factor for power reduction lies in the characteristic time involved.

It is essential to note that, except for traveling wave modulators, an increase in characteristic time always imposes limitations on the signal's bandwidth. The maximum value of characteristic time is approximately T (or slightly less than T by a factor of 2 or 3, which we neglect in our order of magnitude analysis). Then, the maximum reduction of power is $X_{max} \sim (\omega T)^{1/m} \sim (f/B)^{1/m}$, whereas before, $m = \frac{1}{2}$ for EO, AO and second order nonlinearity, and m=1 for the Kerr effect, $f = \omega/2\pi$ is the optical carrier frequency, and $B$ is the signal bandwidth. This maximum power reduction factor that manifests power-bandwidth trade-off is also included in Table 2. The potential becomes apparent when considering electro-optic (EO) modulators featuring finely tuned resonators of high finesse. A plausible outcome is $X_{max} \sim 10^8$ for 20GHz modulation resulting in a reduction of full modulation power to a level below one milliwatt. However, it is crucial to underscore that such outcomes are contingent upon the intricate specifics of implementation, particularly encompassing considerations of loss. In light of this, I shall refrain from further numerical speculation at this juncture.

A few observations can be made – for relatively modest modulation rates one can always use "slow modulators" based on thermal effects carrier injection, but for higher rates, as well as frequency conversion, the EO effect, Kerr effect, and second order nonlinearity are preferable. For ultrafast modulation it is always preferable to use traveling wave schemes.

## 11. Conclusions

Ultimately, this discourse aims to inform the reader on the various complexities, limitations, and trade-offs involved in the design of switches, modulators, frequency converters, and related components by providing a generalized view of the problem. It is definitely not intended to serve as a comprehensive manual guiding designers towards the absolute best implementation. The diverse range of specific demands imposed by different applications, such as the required signal bandwidth, the choice between optical or electrical drive, the importance of a small footprint, tolerance for insertion loss in strong signals, and various other factors, make it impractical to provide a one-size-fits-all solution. Additionally, essential aspects like temperature stability, compatibility with CMOS processes, and cost cannot be overlooked, as they significantly impact the feasibility and success of any design. The paper does not support the notion that a certain modulator made from a particular material will always surpass others; such generalizations would be absurd. Furthermore, this work does not attempt to predict a bright future for specific materials or devices, as there already exists an ample collection of papers making such optimistic claims. Instead, it aims to avoid falling into the realm of futuristic speculation by offering a realistic and grounded perspective. The intention is not to join the ranks of titles that promise revolutionary breakthroughs like "Towards ultra this" or "Towards giant that," which often appear frequently but rarely lead to anything. Having said what this work is not about, let's recap some essential points.

First and foremost, it is important to acknowledge that while routine methods of effective permittivity modulation in the RF domain have been successful, these methods cannot be directly applied to optical frequencies. As a result, attempting to simulate optical phenomena in the RF domain may lead to a remarkable publication[10, 80], but it may have very little practical impact.

Secondly, when it comes to changing the index with minimal power requirements, the most effective approach is to simply replace one material with another or reorient it, as commonly done in liquid crystals. However, these mechanical modulation techniques are limited to low rates, and for applications requiring higher speeds, alternative methods such as EO, AO, or all-optical modulation techniques, which are explored in this study, must be employed

Despite the large variety of index modulation techniques and materials, all of them share a common characteristic: the need to supply a certain amount of energy to induce a large change $\Delta n \sim n$ in the refractive index. This required energy density, termed the VLIC (Very Large Index Change) energy density $u_0$ falls within a similar range of $1-100\ KJ/cm^3$ for all materials and modulation methods, varying only by one or two orders of magnitude, with larger values corresponding to operation in visible in UV regions of spectrum and smaller for IR region.

Due to this consistent energy density requirement, the actual power density (per unit waveguide area) needed to achieve full phase or intensity modulation, as well as frequency conversion, is more dependent on the effective interaction time between the modulating factor (such as electrical, optical, acoustical, or heat wave) and the signal optical wave rather than the specific material being used.

The interaction time can be extended intrinsically through mechanisms such as increased recombination, thermalization, or heat diffusion time in what we refer to as "slow modulation schemes." In these techniques (thermal, photorefractive, modulation via free career excitations) the longer time required to dissipate the modulating energy leads to a larger accumulated index change. Alternatively, interaction time can be prolonged by incorporating resonant structures for either the modulating or signal wave, or by merely increasing the propagation distance.

It is essential to recognize that the enhancement of interaction time is subject to certain constraints, notably the operational speed required for the modulation process. Ideally, the interaction time should be somewhat shorter than the modulation period, but not excessively so. The interaction duration can be modified, for instance, by selecting a material with an appropriate relaxation time or crafting a micro resonator with a suitable Q factor. This balance between interaction time and modulation period ensures effective modulation while meeting the demands of the operational speed.

At the same time, by employing traveling wave modulation, where control and signal waves propagate synchronously, it becomes feasible to decouple the interaction time and modulation speed. As a result, the longer interaction time leads to increased latency, while the modulation speed remains bound by other factors, such as microwave loss. In general, through careful selection of the appropriate "interaction time enhancement" technique (e.g., resonator or traveling wave geometry), it is possible to significantly reduce power consumption for modulation in the 100GHz range and frequency conversion to below a milliwatt in photonic waveguides, and even lower in plasmonic ones. However, it's essential to consider practical implementation factors (footprint, stability, coupling and insertion loss, fabrication cost etc.) that may come into play but are beyond the scope of this paper. Overall, I may express an opinion that most of the improvements in performance of modulators and frequency conversion devices will come from clever designs and improved fabrication rather than from discovery of new materials with extraordinary properties. The reader is free to disagree with this opinion.

In the end , I wish to convey my very guarded optimism that this unified approach to the task of real-time refractive index modification will serve to enlighten the readers, especially students, about the boundaries and trade-offs inherent in designing active optical devices, such as modulators, switches, and frequency conversion devices. Understanding these limitations and considerations can prove instrumental in advancing the development of more power-efficient integrated photonic circuits.

## Acknowledgements

I extend my sincere gratitude to all the pertinent funding sources for their decision not to support this endeavor (and not for the lack of trying on my part). This unique circumstance has afforded me the exceptional privilege to contemplate freely and subsequently articulate my thoughts without excessive restraint. Undoubtedly, the encouragement and enriching discussions with my esteemed colleagues, Prof. P. Noir and Dr. S. Artois, have not only made this endeavor feasible but also immensely gratifying.